\documentclass{aa}  
\usepackage{graphicx}
\usepackage{txfonts}
\begin{document}
\title{Are isolated planetary-mass objects really isolated?}
\subtitle{A brown dwarf-exoplanet system candidate 
in the $\sigma$ Orionis cluster} 
\author{J. A. Caballero\inst{1,2}
     \and
     E. L. Mart\'{\i}n\inst{1,3}
     \and
     P. D. Dobbie\inst{4}
     \and
     D. Barrado y Navascu\'es\inst{5}
}
\offprints{Jos\'e A. Caballero; \email{zvezda@iac.es}}
\institute{Instituto de Astrof\'{\i}sica de Canarias, E-38205 La Laguna,
Tenerife, Spain
\and 
Max-Planck-Institut f\"ur Astronomie, K\"onigstuhl 17, D-69117 Heidelberg, Germany 
\and 
University of Central Florida, Department of Physics, P.O. Box 162385, Orlando, FL
32816-2385, USA  
\and 
Department of Physics and Astronomy, University of Leicester, University Road,
Leicester LE1 7RH, United Kingdom 
\and 
LAEFF-INTA, P.O. Box 50727, E-28080, Madrid, Spain
}
\date{Received 2006 August 2; accepted 2006 August 30}

\abstract
{Free-floating planetary-mass objects have masses below the deuterium burning mass limit at about 13 
Jupiter masses, and have mostly been found in very young open clusters. 
Their origin and relationship to stars and brown dwarfs are still a mystery.}  
{The recent detection by direct imaging of three giant planets at wide separation (50--250 AU) 
from their primaries has raised the question about the 
``true isolation'' of planetary-mass objects in clusters. 
Our goal was to test the possibility that some free-floating planetary-mass object could in fact be 
part of wide planetary systems.}   
{We searched in the literature for stellar and brown-dwarf member candidates of the $\sigma$ Orionis 
cluster ($\sim$ 3\,Ma, $\sim$ 360\,pc) at small angular 
separations from published candidate planetary-mass objects.
We found one candidate planetary system, SE 70,  composed of an X-ray source and a planetary-mass 
object, namely S\,Ori 68, separated by only 4.6 arcsec. 
In order to assess the cluster membership of the X-ray source, we obtained mid-resolution optical 
spectroscopy using ISIS on the William Herschel Telescope.
We also compiled additional data on the target from available astronomical catalogues.}  
{We have found that SE 70 follows the spectrophotometric sequence of the cluster and displays 
spectroscopic features of youth, such as lithium in absorption 
and chromospheric H$\alpha$ emission.
The radial velocity is consistent with cluster membership. 
Hence, SE 70 is very probably a $\sigma$ Orionis cluster member. 
The projected physical separation between SE 70 and S\,Ori 68 is 1\,700$\pm$300\,AU at the distance of the cluster. 
If the common proper motion is confirmed in the near future, the system would be
composed of an M5--6 
brown dwarf with an estimated mass of $\sim$ 45\,M$_{\rm Jup}$ 
and an L5$\pm$2 giant planet with an estimated mass of $\sim$ 5\,M$_{\rm Jup}$.
It would be the widest and one of the lowest-mass planetary systems known so far.} 
{}
\maketitle
%

\section{Introduction}

Planetary-mass objects (PMOs) were {\em directly} detected for the first time
in the $\sigma$ Orionis cluster and in other very young star-forming regions 
(Zapatero Osorio et al.\ 2000; Lucas \& Roche 2000; Najita et al.\ 2000).
Recent discoveries suggest that PMOs are abundant in clusters and include
the detection of objects with masses potentially as low as $\sim$ 3 Jupiter masses (M$_{\rm Jup}$) (Zapatero Osorio et al.\ 2002c; Lucas et al.\ 2005).
Furthermore, observations suggest that many PMOs are surrounded by
discs (Luhman et al.\ 2005).
Although cluster PMOs have masses below the 
deuterium-burning mass limit ($\sim$ 13\,M$_{\rm Jup}$; Chabrier \& Baraffe 2000), they have not achieved 
the ``grade'' of exoplanet because they appear to float freely in the intracluster 
medium, far from the gravitational link to other star cluster members.
The direct detection of planetary-mass candidates around a brown dwarf of the
TW Hydrae association (Chauvin et al.\ 2004 -- at a projected physical 
separation of 55\,AU) and around the stars \object{AB Pictoris} and \object{GQ Lupi} 
(Chauvin et al.\ 2005 -- 250\,AU; Neuh\"auser et al.\ 2005 -- 100\,AU) 
presents the opportunity to photometrically and spectroscopically characterize exoplanets at orbital separations much larger than those indirectly detected 
with radial velocity surveys (the radial velocity exoplanet candidate with the 
largest semi-major axis is 55 Cnc d, with $a$ = 5.3$\pm$0.9\,AU; Marcy et al. 2002).
The discussion on how PMOs in wide orbits can be formed has been fed by the recent discovery of the system \object{Oph 162225--240515} AB (240\,AU) in the 1\,Ma-old Ophiuchus region, which may be a very low-mass brown dwarf and a PMO pair (Jayawardhana \& Ivanov 2006) or a low-mass brown-dwarf binary (Close et al. submitted).

Here, we present the first results of a survey for wide planetary systems
in the $\sigma$ Orionis cluster. 
We have searched for candidate low-mass stellar and brown-dwarf members in close proximity to known PMOs. 
Our knowledge of the stellar and brown-dwarf populations of the cluster has dramatically increased in the last 
few years (B\'ejar et al.\ 1999, 2001, 
2004 -- mostly focused on very low mass stars and brown dwarfs; Scholz \& Eisl\"offel 2004; 
Caballero et al.\ 2004; Sherry et al.\ 2004; 
Kenyon et al.\ 2005; Burningham et al.\ 2005; Hern\'andez et al.\ 2005).  
Prior to these studies, only massive and/or notably active stars were
known in the cluster (e.g.\ Haro \& Moreno 1953; Warren \& Hesser 1978;
Wiramihardja et al.\ 1989; Brown et al.\ 1994; Wolk et al.\ 1996). 
This cluster has become a favourite hunting ground for brown dwarfs and PMOs because it is 
nearby (360$^{+70}_{-60}$ pc), young (3$^{+5}_{-2}$ Ma) 
and free of extinction (e.g.\ Caballero 2005; Jeffries et al.\ 2006).

\subsection{SE 70 and S\,Ori 68}

The search for cluster candidate members at projected separations less than 15 arcsec (about 5\,000\,AU at the cluster distance) from isolated PMO candidates of $\sigma$ Orionis had been unfruitful until the recent spatial and spectral analysis of a full {\em XMM-Newton}/EPIC field in the cluster performed by 
Franciosini, Pallavicini \& Sanz-Forcada (2006).
In that work, out of the 175 X-ray sources detected, 88 were identified with cluster
member candidates extracted from the literature.
In two cases, the identification was ambiguous since there were two cluster
member candidates falling inside their identification radii.
One of these systems, associated with the X-ray source NX 67, 
was formed by an intermediate M-type brown-dwarf candidate proposed 
by Scholz \& Eisl\"offel (2004), named SE 70, and by a PMO candidate, S\,Ori 68.
NX 67 underwent a flare during the observation, with an increase in the count rate 
over the quiescent level by a factor of $\sim$ 4 and a total duration of about 3\,h.
Although Franciosini et al.\ (2006) did not exclude the presence of a very weak
contribution of S\,Ori 68 to the total X-ray flux, they concluded that the
bulk of it was associated with SE 70.

The planetary-mass object candidate S\,Ori 68 ($\sim$ 5\,M$_{\rm Jup}$), 
previously classified as ``isolated'', is about 4.6 arcsec southeast of 
SE 70 ($\theta \approx$ 140\,deg).
At the distance of the $\sigma$ Orionis cluster, the projected physical separation 
is 1\,700$\pm$300\,AU.
S\,Ori 68 was firstly identified by Zapatero Osorio et al.\ (2000) as a 
free-floating planet candidate from an $IJ$ survey with follow-up in the $K$ band.
Barrado y Navascu\'es et al.\ (2001) classified it as an L5.0$\pm$2.0-type
object from an optical spectrum taken with FORS at the Very Large Telescope.
The low resolution and the insufficient signal-to-noise ratio of the data did 
not allow them to detect spectroscopic features of extreme youth 
(i.e.\ Li {\sc i} $\lambda$6707.8 in absorption, strong asymmetric H$\alpha$ emission, 
anomalous broadening of atomic absorption lines due to low gravity).
Although the object follows the spectrophotometric cluster sequence, 
more data are needed to confirm its membership in the $\sigma$ Orionis 
cluster. 

Nevertheless, the proximity between SE 70 and S\,Ori 68 motivated us to characterize in more detail the physical properties of the primary. 
We have taken intermediate-resolution optical spectroscopy and have compiled astrometric and photometric data of SE 70 to test more rigorously the hypothesis of membership in the $\sigma$ Orionis cluster.

\section{Observations and data mining}

\subsection{Optical spectroscopy}

\begin{figure}
\centering
\includegraphics[width=0.50\textwidth,angle=0]{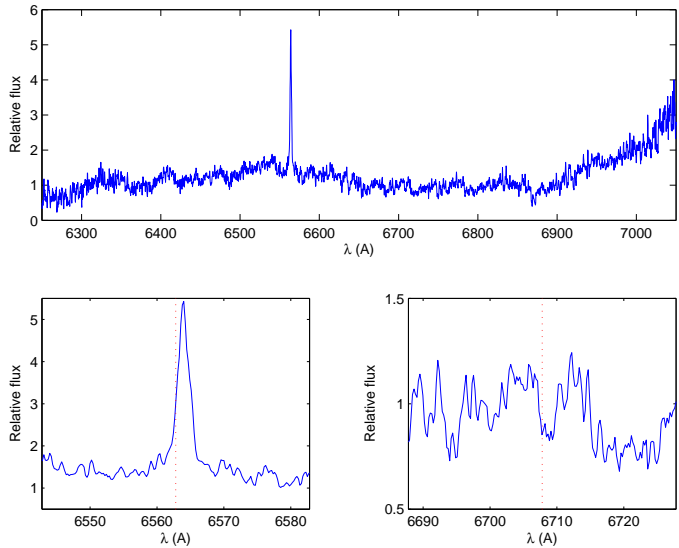}
\caption{Final combined spectrum of SE 70, taken with ISIS/WHT (F$_\lambda$, 3-pixel smoothing).
Top window: whole spectral range.
Bottom windows: portions of 20\AA~width centred in H$\alpha$ (left) and Li {\sc i} (right).
Line identifications are marked with dotted vertical lines.
The spectrum has not been corrected for heliocentric or intrinsic radial velocities.}
\label{fe02}
\end{figure}

On 2006 February 4, we used the red arm of ISIS at the 
4.2\,m William Herschel Telescope (WHT) at the Observatorio del Roque de los 
Muchachos to obtain four mid-resolution spectra of SE 70.
ISIS is a high-efficiency, double-armed, medium-resolution 
(8--120\,\AA\,mm$^{-1}$) spectrograph. 
The detector mounted on the red arm of ISIS is a Marconi2 with a 4610 pixel axis along the dispersion direction.
The four spectra covered the wavelenght range 6\,200--7\,100\,\AA~(central $\lambda$ = 6561\,\AA), which  
allowed us to study 
the H$\alpha$ and Li {\sc i} $\lambda$6707.8 spectroscopic features, and also to estimate both the spectral 
type and the radial velocity of the object.
Nominal dispersion with the red grating with 1\,200 rulings\,mm$^{-1}$ (R1200R) was 16.6\,\AA\,mm$^{-1}$, which 
together with the 
1.0-m slit led to a spectral resolution  $R \sim$ 3\,000.
The seeing was good (0.6--0.8\,arcsec) and the sky was fairly clean.
The total exposure time of the final co-added spectrum was 4 $\times$ 1\,800\,s = 2\,h.
The spectra reduction was typical, with bias subtraction, flat correction, wavelength calibration 
(with arc spectra taken at the 
same position of the sky as the target) and flux normalization (we observed one flux standard star
 -- \object{HD 289002}, B3V -- just before and after SE 70).
The whole wavelength interval of the final combined spectrum of SE 70 is shown in the top window of Figure \ref{fe02}.

We estimated the spectral type of SE 70 at M5$\pm$1 from the calculation of the following spectral indices: 
$PC1$, $I1$, $I2$, $I3$ (Mart\'{\i}n, 
Rebolo \& Zapatero Osorio 1996; Mart\'{\i}n \& Kun 1996). 
All of them are based on the the ratio between the spectral fluxes at two different wavelengths, which roughly 
provides the slope of the spectral energy 
distribution in the red optical range. 
Calculated indices from the co-added SE 70 spectrum were compared to a grid of indices for different spectral types (early to late M) and luminosity classes (V to III). 
The dwarf class provided the best fit.
The spectral types determined for each index were in the range M4.0 to M6.5, M5.0 being  the 
most probable spectral type.
This value is between 1.0 and 1.5 spectral subtypes earlier than expected from the $J$-band magnitude of SE 70 for a cluster 
member (see next section), but is consistent within the 
error bars.
It should be noted that it is difficult to tightly constrain the spectral
type from such a short wavelength interval, which 
allows us to use only a limited number of spectral type indicators.
The strength of the titanium and vanadium oxide bands and the increasing slope of the spectrum 
redwards of 6\,900\,\AA~provide further proofs of the 
relatively late spectral type of SE 70.

   \begin{table}
      \caption[]{Spectroscopic data of SE 70 and S\,Ori 68.}
         \label{spectroscopicdata}
     $$ 
         \begin{tabular}{l c c}
            \hline
            \hline
            \noalign{\smallskip}
 		& SE 70$^{a}$ 			& S\,Ori 68$^{b}$ 	\\ 
            \noalign{\smallskip}
            \hline
            \noalign{\smallskip}
Sp. type 		& M5.0$\pm$1.0		& L5.0$\pm$2.0 \\ 
pEW(H$\alpha$) (\AA) 	& --4.5$\pm$1.0 	& $<$ 20 \\ 
pEW(Li {\sc i}) (\AA) 	& +0.60$\pm$0.10 	& --- \\ 
V$_r$ (km\,s$^{-1}$) 	& +40$\pm$10 		& --- \\ 
            \noalign{\smallskip}
            \hline
         \end{tabular}
     $$ 
\begin{list}{}{}
\item[$^{a}$] SE 70 (IAU name: \object{[SE2004] 70}; also: NX 67, 2MASS J053838.88--022801.6, DENIS J053838.9--02280, USNO-B1.0 0875--0103266).
\item[$^{b}$] S\,Ori 68 (IAU names: \object{[BZR99] S\,Ori 68}, [BMZ2001] S\,Ori J053839.1-022805).
Spectral type and H$\alpha$ pseudoequivalent width upper limit from Barrado y Navascu\'es et al.\ (2001). 
\end{list}
   \end{table}

The H$\alpha$ $\lambda$6562.8 atomic feature is in emission.
{\em All} the M-type stars and brown dwarfs in the $\sigma$ Orionis cluster with available spectroscopy display H$\alpha$ 
in emission (see, for example, Zapatero Osorio et al.\ 2002a; Caballero 2006 -- this fact is also applicable to all the K-type stars of the cluster).
In general, this emission is due to chromospheric activity, and the spectra show pseudo-equivalent widths of the 
H$\alpha$ line, pEW(H$\alpha$), of about --5\,\AA.
However, in a few cases, the emission is due to accretion from a disc, with pEW(H$\alpha$) of a few tens to a  
few hundred \AA~(Zapatero 
Osorio et al.\ 2002a, 2002b; Caballero et al.\ 2006).
We have measured the pEW(H$\alpha$) in the final combined spectrum of SE 70 at $-4.5 \pm 1.0$\,\AA.
According to the criterion defined by Barrado y Navascu\'es \& Mart\'{\i}n (2003), SE 70 is not an accretor.
The pEW(H$\alpha$)s in each of the four individual spectra range between 3.6 and 5.4\,\AA, which are consistent 
with the stability of the emission at the 1 $\sigma$ level.
The profile of the H$\alpha$ emission is shown in the bottom left window of Figure \ref{fe02}.
The signal-to-noise ratio of the final spectrum does not allow us to detect any asymmetry.

We have also detected Li {\sc i} $\lambda$6707.8 in absorption, with a pseudo-equivalent width, pEW(Li {\sc i}), of +0.60$\pm$0.10\,\AA.
This value is similar to those of other members of the $\sigma$ Orionis cluster with intermediate M spectral type (Zapatero Osorio et al.\ 2002a; Kenyon et al.\ 2005). 
Since objects of this spectral type destroy all their lithium through nuclear reactions in only a few megayears, it is derived that SE 70 is an extremely young M-type object with an age between 1 and $\sim$ 10\,Ma (the minimum and maximum ages for the $\sigma$ Orionis cluster are 1 and 8\,Ma, respectively).

Using the atomic features that are visible in the wavelength interval studied, H$\alpha$, Li {\sc i} 
and some Fe {\sc i} lines, and the heads of some 
titanium oxide bands, we have estimated the radial velocity of SE 70 at about 40\,km\,s$^{-1}$ (the 
heliocentric velocity correction for the mean 
time of observation was $-$25.3\,km\,s$^{-1}$).
If, assuming a conservative uncertainty of 25\,\%, 
the estimated value is consistent with other more precise determinations of the mean radial velocity 
of the cluster found in the literature 
(at about 30--35\,km\,s$^{-1}$ -- Walter et al.\ 1998; Zapatero Osorio et al.\ 2002a; Muzerolle et 
al.\ 2003; Kenyon et al.\ 2005; 
Burningham et al.\ 2005; Caballero 2006; Jeffries et al.\ 2006 --group 1).
H$\alpha$ emission, if not chromospheric, could be blue-/redshifted, which would introduce an additional 
source of uncertainty into our measurement.
We have also determined the radial velocity of SE 70 with respect to several emission lines that may be 
adscribed to the nebulosity in the line of sight, and that
are probably associated with the star-forming region.
The difference between both radial velocities is null with an accuracy of 5\,km\,s$^{-1}$.

In Table \ref{spectroscopicdata} we summarize the measurements mentioned above (spectral type, H$\alpha$ 
and Li {\sc i} pseudo-equivalent widths and radial velocity of SE 70).
For completeness, the spectral type and H$\alpha$ pseudo-equivalent width upper limit from Barrado y 
Navascu\'es et al.\ (2001) of S\,Ori 68 are also provided. 
Spectra with higher signal-to-noise ratio for both objects would be desirable for confirming membership 
to the $\sigma$ Orionis cluster.

\subsection{Additional data}

\begin{figure}
\centering
\includegraphics[width=0.49\textwidth,angle=0]{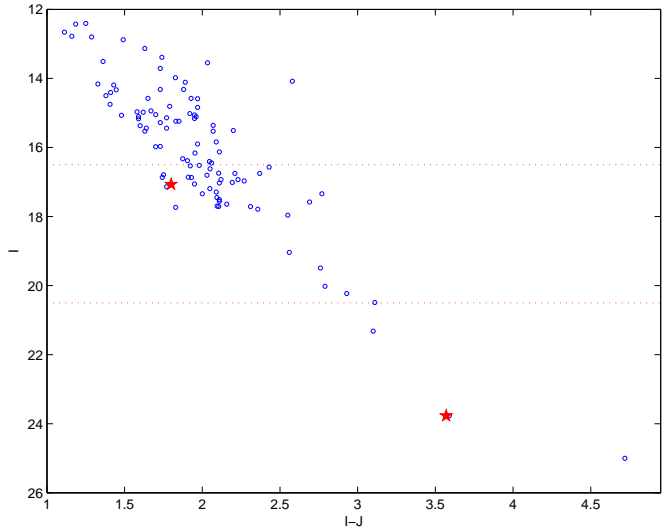}
\caption{$I$ vs.\ $I-J$ colour--magnitude diagram showing the $\sigma$ Orionis members fainter 
than $I$ = 12\,mag with spectroscopic features of youth compiled by Caballero (2005), in small 
open circles, and SE 70 and S\,Ori 68, in large filled stars.
Rough hydrogen (top) and deuterium (bottom) burning mass limits are shown with dotted lines.} 
\label{fe05}
\end{figure}

The coordinates of the optical and near-infrared counterparts of SE 70 in
the 2MASS, DENIS and USNO-B1.0 catalogues match between them with a precision
better than 0.\,1arcsec. 
Coordinates of this target in the discovery paper by Scholz \& Eisl\"offel
(2004) do not deviate from those of 2MASS by more than 0.5\,arcsec.
Those authors did not reported SE 70 to be a photometric variable in the $I$ band
and estimated an upper limit of 0.035\,mag for its photometric stability. 
They announced an optical magnitude of $I$ = 16.60\,mag, which led to a comparatively blue $I-J_{\rm 2MASS}$ colour for its magnitude if it is cluster member. 
However, based on $I$-band data from DENIS catalogue, we determine an $I-J$ colour of 1.80$\pm$0.13 mag, which locates this source on the the bluest part of the spectrophotometric cluster sequence in the $I$ vs.\ $I-J$ colour--magnitude
diagram, in the region of other M5--7 cluster members with lithium in absorption (Figure \ref{fe05}).
SE 70 was not selected as a photometric candidate during other optical searches in the area (B\'ejar 
et al.\ 1999, 2001; Kenyon et al.\ 2005).
Some bright brown-dwarf cluster members with proven membership, such as the classical T Tauri substellar 
analogue \object{S\,Ori J053825.4--024241}, 
display quite blue colours in the red optical and had also escaped  previous searches (Caballero et 
al.\ 2006).

   \begin{table}
      \caption[]{Astrometric and photometric data of SE 70 and S\,Ori 68}
         \label{basicdata}
     $$ 
         \begin{tabular}{l c c}
            \hline
            \hline
            \noalign{\smallskip}
 		& SE 70$^{a}$ 			& S\,Ori 68$^{b}$ 	\\ 
            \noalign{\smallskip}
            \hline
            \noalign{\smallskip}
$\alpha$		& 05 38 38.89		& 05 38 39.1 	\\   
$\delta$		& --02 28 01.6		& --02 28 05	\\ 
$\mu_\alpha \cos{\delta}$ (mas\,a$^{-1}$)& --14$\pm$7	& --- \\
$\mu_\delta$ (mas\,a$^{-1}$)& +8$\pm$3		& --- \\
            \noalign{\smallskip}
            \hline
            \noalign{\smallskip}
$B$			& 20.68$\pm$0.56	& --- \\ 
$R$			& 19.05$\pm$0.24	& --- \\ 
$I$  			& 17.07$\pm$0.12	& 23.77$\pm$0.17 \\ 
$J$ 			& 15.27$\pm$0.04	& 20.2$\pm$0.3 \\ 
$H$			& 14.73$\pm$0.05	& --- \\ 
$K_{\rm s}$ 		& 14.43$\pm$0.07	& 18.4$\pm$0.3 \\ 
$[3.6]$ 		& 14.06$\pm$0.14	& --- \\ 
$[4.5]$ 		& 14.07$\pm$0.17	& --- \\ 
$[5.8]$ 		& 13.76$\pm$0.19	& --- \\ 
$[8.0]$ 		& 13.8$\pm$0.3		& --- \\  
            \noalign{\smallskip}
            \hline
         \end{tabular}
     $$ 
\begin{list}{}{}
\item[$^{a}$] 
Coordinates $\alpha$ and $\delta$ (J2000) and $JHK_{\rm s}$ magnitudes from 2MASS (Cutri et al.\ 2003); 
$B$ (averaging two detections) and $R$ magnitudes (averaging two double
detections) and tangential velocities $\mu_\alpha$ and $\mu_\delta$ from
USNO-B1.0 catalogue (Monet et al.\ 2003);  
$I$ magnitude from DENIS catalogue (The DENIS Consortium); 
{\em Spitzer}-band magnitudes from Caballero et al.\ (in prep.).
\item[$^{b}$] 
Coordinates $\alpha$ and $\delta$ (J2000) and $IJK$ magnitudes
from Zapatero Osorio et al.\ (2000). 
\end{list}
   \end{table}

\begin{figure}
\centering
\includegraphics[width=0.49\textwidth,angle=0]{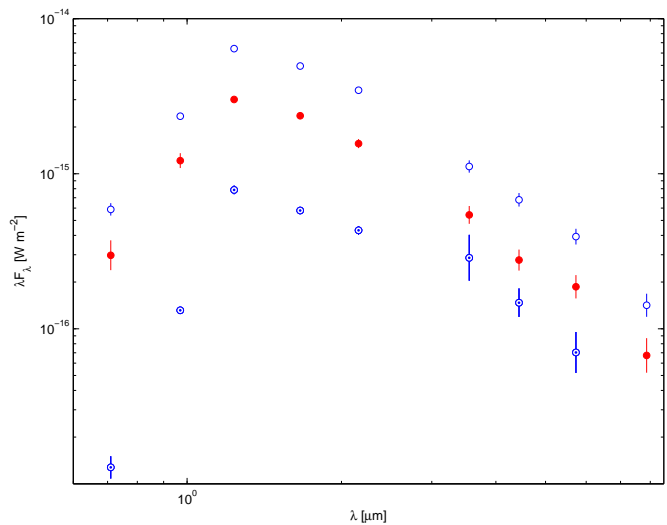}
\caption{Spectral energy distributions of SE 70 from the optical $R$ band to 8.0
$\mu$m (filled circles, `$\bullet$')
The spectral energy distributions of two very low-mass cluster members with lithium in absorption and other cluster membership indices are also given for comparison.
They are \object{S\,Ori J053911.4--023333}, M5.0$\pm$0.5 (open circles `$\circ$'), and \object{S\,Ori 45}, M8.5$\pm$0.5 (solar symbol, `$\odot$').} 
\label{fe03}
\end{figure}

Additional photographic $B$- and $R$-band and $[3.6]$-, $[4.5]$-, $[5.8]$- and $[8.0]$-band photometry has been taken from the USNO-B1.0 catalogue 
and the IRAC/{\em Spitzer} Telescope archive, respectively (see Caballero et al., in prep., for the latter).
In Table \ref{basicdata} we provide all the collected photometry of SE 70.
The extremely red optical--infrared colours of the target support the late-M
spectral type derived spectroscopically.
Furthermore, the spectral energy distribution, shown in Figure \ref{fe03}, is identical to those of other non-accreting M5--7 very low-mass 
stars and brown dwarfs of the $\sigma$ Orionis cluster.
Figure \ref{fe03} also shows the spectral energy distribution of a slightly warmer object in the star-brown dwarf boundary (M5.0$\pm$0.5) and a cooler brown dwarf (M8.5$\pm$0.5).
Both objects display spectral features of extreme youth and are confirmed cluster members (Zapatero Osorio et al. 2002, 2003; Kenyon et al. 2005).
There is no evidence of any surrounding disc at wavelengths longer than 2.0\,$\mu$m from the shape of the spectral energy distribution of SE 70.
Redwards of 1\,$\mu$m, it provides a reasonable fit to   that of a black body at 360\,pc with 
$T_{\rm eff}$ = 3\,200\,K, $R$ = 0.42\,R$_\odot$ (bluewards 
of 1\,$\mu$m there are strong water and metallic oxide bands that affect the continuum).
The estimated effective temperature is consistent with the M5--6 spectral type determination. 

Table \ref{basicdata} also provides the astrometry of SE 70.
Coordinates indicate that it is located at about 8\,arcmin to the O9.5V star that gives the name to the cluster, 
\object{$\sigma$ Ori A} (and 
which is roughly the centre of mass of the cluster).
Tangential velocities in $\alpha$ and $\delta$ have been taken  from the USNO-B1.0 catalogue.
The difference between the proper motions of SE 70 and $\sigma$ Ori A is as large as 
($\Delta \mu_\alpha$,$\Delta \mu_\delta$) = (--19$\pm$7,+8$\pm$3)\,mas\,a$^{-1}$.
These values are several times larger than the cluster velocity dispersion.
The automatic proper motion determination at the faint optical magnitudes of SE 70, fainter than 
the completeness $BR$-band magnitudes of the USNO-B1.0 
catalogue, may lead to unreliable tangential velocities or to underestimated error bars.
This is probably the case of SE 70.
A simple inspection of the tangential velocity values of several faint spectroscopically confirmed 
members of the $\sigma$ Orionis cluster supports this 
hypothesis (they also display proper motions that are abnormally different from zero).

\section{Discussion}

Except for the unreliable astrometric measurement, all the studied astrometric parameters indicate
 that SE 70 is a bona fide non-accreting 
$\sigma$ Orionis cluster member: it displays X-ray emission, chromospheric H$\alpha$ emission, Li {\sc i} 
in absorption and radial velocity similar to 
those of the cluster and of the nebulosity, and lies on the spectrophotometric cluster sequence.
The derived spectral type is on the borderline between very low-mass stars and brown dwarfs in $\sigma$ Orionis.
However, the $J$-band magnitude of SE 70, which marks the peak in the spectral energy distribution, is about 
1.0\,mag fainter than those of the most 
masive brown dwarfs in the cluster (at the hydrogen-burning mass limit, $\sim$ 75\,M$_{\rm Jup}$; B\'ejar 
et al.\ 2001; Caballero et al.\ 2004).
Following our procedure for mass determination in the substellar domain, detailed in Caballero (2006)\footnote{The mass is derived from the comparison between theoretical luminosities of Chabrier et al.\ (2000) and the luminosities of the objects.
The latter are computed from the $J$-band absolute magnitude and an $I-J$ colour-dependent bolometric correction.}, we estimate SE 70 to be a brown dwarf with a
mass of 45\,M$_{\rm Jup}$.
Hence, S\,Ori 68 and SE 70 may form a system whose masses resemble those of the Chauvin et al.\ (2004) \object{2MASS J1207334--393254} system (about 45 + 5\,M$_{\rm Jup}$ in the case of SE 70 + S\,Ori 68 versus about 25 + 5\,M$_{\rm Jup}$ in the case of 2M1207--39 + 2M1207--39b).
However, the projected physical separation of the hypothetical $\sigma$ Orionis planetary system is about 30 times larger than that of 2M1207--39, and 
about seven times larger than that of AB Pic and Oph 162225--240515.
Due to the low mass and the wide separation of the SE 70 + S\,Ori 68 system, it has the lowest gravitational binding energy among the considered planetary systems.
Given the relatively high spatial density of cluster members in $\sigma$ Orionis and the weak bounding between SE 70 and S\,Ori 68, it seems dubious that the system can survive the tidal field within the cluster.
The existence of a binary system with these characteristics is evidently a challenge for present ultra low-mass star forming scenarios, which predict tighter and more massive binary/planetary systems. 

\begin{figure}
\centering
\includegraphics[width=0.49\textwidth,angle=0]{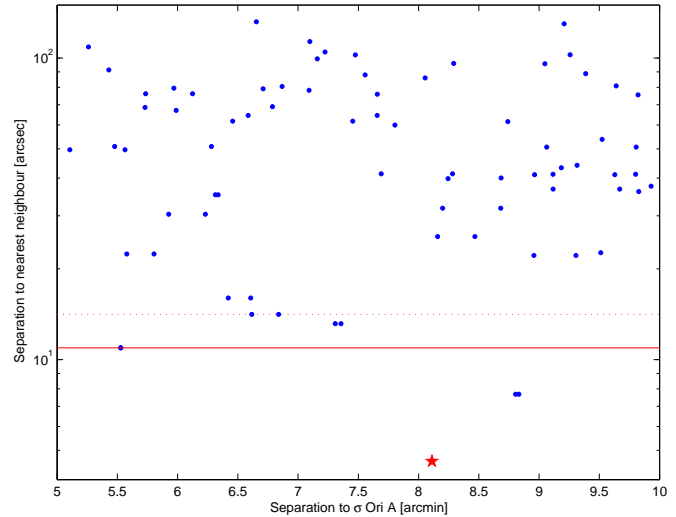}
\caption{Angular separation to the nearest neighbour vs.\ separation to the centre of the $\sigma$ Orionis cluster.
Small dots represent the star and massive brown-dwarf photometric member candidates in the study by Caballero 
(in prep.), while the large 
filled stars represent the SE 70 + S\,Ori 68 system.
Low-mass brown-dwarf and IPMO cluster member candidates are not included.
Horizontal lines denote the 90 (dotted) and 95\,\%
(solid) confidence limits below which the probabilities for a pair at 5 to 10\,arcmin to $\sigma$ Ori A to 
be a visual alineation are less than 10 and 5\,\%,
respectively.
There are only two systems below the 95\,\%
confidence limit: [W96] 4771--899 AB + C and SE 70 + S\,Ori 68.
The results from 0 to 30\,arcmin to the cluster centre and the description of the spatial distribution study 
will be given in Caballero (in prep.). 
} 
\label{fe04}
\end{figure}

There is the possibility of the pair being a visual projection of two unrelated cluster members at 
slightly different heliocentric distances 
and hence not being a truly binary system.
This is in contradiction with the statistical analysis performed by Caballero (in prep.),
who has studied the spatial distribution of {\em all} the stellar population and the highest-mass 
domain of the brown-dwarf population in the $\sigma$ Orionis cluster.
He has shown that 95\,\% 
of the studied cluster member candidates at angular separations to $\sigma$ Ori A between 5 and 
10 arcmin have their nearest neighbourhood at 
more than 10.9\,arcsec (the SE 70 + S\,Ori 68 system is at about 8\,arcmin to the cluster centre).
Among the more than 400 cluster member candidates studied  from a 2MASS + DENIS correlation, none 
has a nearest neighbour closer than 7.1\,arcsec, 
while the separation between SE 70 and S\,Ori 68 is only 4.6\,arcsec.
Figure \ref{fe04} illustrates this discussion, and indicates that the SE 70 + S\,Ori 68 pair has 
less than about 1\,\% 
probability to be a chance visual alignment.
There are stellar pairs in the $\sigma$ Orionis cluster with smaller separations between components, 
but they are spectroscopic binaries 
(e.g.\ \object{OriNTT 429}; Lee et al.\ 1994) or
very tight binaries unresolved by 2MASS or DENIS (e.g.\ $\sigma$ Ori A and 
\object{$\sigma$ Ori B} -- 0.250\,arcsec; $\sigma$ Ori AB and 
\object{$\sigma$ Ori IRS1} -- 3.32$\pm$0.15\,arcsec; \object{HD 37525 A} 
and \object{HD 37525 B} -- 0.47$\pm$0.04\,arcsec; \object{[W96] 4771--899 A} 
and \object{[W96] 4771--899 B} -- 0.40$\pm$0.08\,arcsec; Caballero 2005)\footnote{The 
tight pair [W96] 4771--899 AB and \object{[W96] 4771--899 C} 
(aka \object{S\,Ori J053847.7--022711}, at $\rho$ = 7.63$\pm$0.10\,arcsec) is in fact a 
triple stellar system.
There is an additional binary system candidate formed by a very low-mass star, S\,Ori 
J0539268--026614, and a faint brown dwarf at 4.4\,arcsec 
($J$ = 16.21$\pm$0.09\,mag, not found in 2MASS or DENIS catalogues).
Both of them show Li {\sc i} in absorption (Zapatero Osorio, B\'ejar et al. in prep.).}.

It is not the aim of this paper to prove beyond doubt the cluster membership of the planetary-mass candidate S\,Ori 68.
In order to do that, high-quality mid-resolution near-infrared spectra of S\,Ori 68 should be taken to detect 
low-gravity spectral features in the 
absorption atomic lines of alkalis, such as sodium or potassium.
These features have been already found in several spectra of young brown dwarfs (e.g.\ Gorlova et al.\ 2003).
To date, this study may only be accomplished, and with a time-consuming effort, with instruments such as LRIS at the Keck Observatory.
Given the extraordinary faintness of S\,Ori 68 in the red optical range ($I \sim$ 24\,mag, $R >$ 26--27\,mag; S\,Ori 68 has the same $I$- and $J$-band magnitudes as \object{S\,Ori 66}, an intermediate-L planetary-mass object in $\sigma$ Orionis with H$\alpha$ in moderate emission), the detection of lithium in absorption in an 
optical spectrum is practically out of the question with present instrumentation.

Further ultra-accurate proper motion studies are needed to verify that both S\,Ori 68 and SE 70 form a common proper motion system.
The Ori OB1b Association is in the solar antapex, moving in the opposite direction with respect to the Sun.
Hence, the tangential velocities of the stellar members of the association are very low,  less than 
5\,mas\,a$^{-1}$ according to de Zeeuw et al.\ (1999).
Confirmation of the common proper motion would require an accuracy of a few hundred milliarsec.
Such accurate measurements on very faint objects only be will achievable with technology of the near future, such as the 
Large Binocular Telescope and the LINC-NIRVANA instrument.
It could be also done with the {\em Hubble} Space Telescope if  a large enough time baseline is used.
However, it could not be done with the {\em GAIA} satellite, because of faintness of targets in the optical.

Radial velocity measurements of the pair obtained with current or future near-infrared high-resolution spectrographs attached to $\ge$10\,m-class telescopes, like NIRSPEC/Keck or NAHUAL/Gran Telescopio Canarias (Mart\'{\i}n et al. 2005), might be able to constrain the binding status of the SE 70 $+$ S\,Ori 68 system.
These measurements must be more precise than 1\,km\,s$^{-1}$ to disentangle the common proper motion of the hypothetical binary system and the velocity dispersion of the $\sigma$ Orionis cluster (that is of the order of 2.4\,km\,s$^{-1}$ -- Caballero 2005).
To illustrate the capabilities of NIRSPEC/Keck, Mart\'{\i}n et al. (2006) reached a precision of 360\,m\,s$^{-1}$ in the radial velocity determination of the M9V-type brown dwarf \object{LP 944--20}.

\section{Summary}

We report on the analysis of a probable a brown dwarf--exoplanet system candidate in the $\sigma$ Orionis cluster, formed by SE 70 and S\,Ori 68.
The projected physical separation between them is 1\,700$\pm$300\,AU at the distance to the cluster.
We have derived the spectral type of the primary (in the M5--6 interval) and found spectroscopic features typical in other young members of the $\sigma$ Orionis cluster: 
lithium in absorption, H$\alpha$ in (probably chromospheric) emission and radial velocity similar to the mean radial velocity of the cluster.
Franciosini et al.\ (2006) had previously found the primary to be an X-ray emitter that underwent a flare during their observations.
From an analysis of the spatial distribution of the cluster, if S\,Ori 68 is also a cluster member, there is about a 99\,\% 
probability that they form a real planetary system.
If this is the case, not all ``isolated'' planetary-mass objects are really isolated, and some planetary systems may have separations as wide as those of binary stellar systems.

\begin{acknowledgements}

We thank V\'{\i}ctor J. S. B\'ejar, Rafael Rebolo and Mar\'{\i}a Rosa Zapatero Osorio for helpful comments.  
We also thank M. F. Sterzik for suggestions that have improved the quality of the manuscript and for his fast refereeing.
Partial financial support was provided by the Spanish Ministerio de Ciencia y 
Tecnolog\'{\i}a project AYA2005--06453 of the Plan Nacional de 
Astronom\'{\i}a y Astrof\'{\i}sica and by NSF grant AST0440520. 
Based on observations made with the William Herschel Telescope operated
on the island of La Palma by the Isaac Newton Group in the Spanish
Observatorio del Roque de Los Muchachos of the Instituto de Astrof\'{\i}sica
de Canarias. 
This work is based in part on observations made with the Spitzer Space Telescope, which is operated by the Jet Propulsion Laboratory, California Institute of Technology under a contract with NASA.

\end{acknowledgements}

\end{document}